\documentclass[3p,times,twocolumn]{elsarticle}

\usepackage{ecrc}

\usepackage{slashed}
\usepackage{subcaption}
\usepackage{adjustbox}
\usepackage{mdframed}
\usepackage{amsmath}

\volume{00}

\firstpage{1}

\journalname{Nuclear and Particle Physics Proceedings}

\runauth{S. Hauksson et al.}

\jid{nppp}

\jnltitlelogo{Nuclear and Particle Physics Proceedings}

\usepackage{amssymb}

\usepackage[figuresright]{rotating}

\begin{document}

\begin{frontmatter}



\dochead{}

\title{Bulk viscous corrections to photon production in the quark-gluon plasma}


\author[McGill]{Sigtryggur Hauksson}
\author[Brookhaven]{Chun Shen}
\author[McGill]{Sangyong Jeon}
\author[McGill]{Charles Gale}

\address[McGill]{Department of Physics, McGill University, 3600 University Street, Montreal, Quebec, H3A 2T8, Canada}
\address[Brookhaven]{Physics Department, Brookhaven National Laboratory, Upton, NY 11973, USA}

\begin{abstract}
Photons radiated in heavy-ion collisions are a penetrating probe, and as such can play an important role in the determination of the quark-gluon plasma (QGP) transport coefficients. 
In this work we calculate the bulk viscous correction to photon production in two-to-two scattering reactions. 
Phase-space integrals describing the bulk viscous correction are evaluated explicitly in order to avoid the forward scattering approximation which is shown to be poor for photons at lower energies. We furthermore present hydrodynamical simulations of AA collisions focusing on the effect of this calculation on photonic observables. Bulk corrections are shown to reduce the elliptic flow of photons at higher \(p_T\).

\end{abstract}


\end{frontmatter}


\section{Introduction}
\label{Introduction}

Photons are an important probe in relativistic heavy-ion collisions: they are created at all stages of the collisions and leave the medium undistorted by the strong interaction. At leading order in perturbation theory, ${\cal O}(\alpha_s)$, photon production in the QGP phase receives contributions from 
  two-to-two scattering channels \cite{Baier1991, Kapusta1991} which comprise Compton scattering and quark-antiquark annihilation, and from  channels which consider quark bremsstrahlung, off-shell pair annihilation, and coherence between different scattering sites (Landau-Pomeranchuk-Migdal effect) \cite{AMY2001}. The early works assumed a medium in thermal equilibrium.  However, hydrodynamical simulations show that there are sizeable viscous effects in the evolution of the created medium due to shear and bulk viscosity \cite{Ryu2015}. For theoretical consistency, one  thus needs to include the effect of viscosity on photon production. In return, this allows for the extraction of QGP transport coefficients from electromagnetic observables: a worthy reward. 


Table II in Ref. \cite{Paquet2015} presents a summary of the viscous corrections (which include both shear and bulk)  to photon production included so far in photon calculations. The contribution presented here addresses the correction owing to bulk viscosity in a specific channel:  two-to-two scattering channels, for which estimates of the bulk viscous correction existed using the forward scattering approximation (see below).  We present the results of a field-theoretical calculation, and then explore its phenomenological effects using relativistic hydrodynamics.


\vspace*{-0.2cm}
\section{Bulk viscous correction to two-to-two channels}

The rate of photon emission, \(R\), in a unit volume can be written in a way which is valid out of equilibrium, as long as the medium is static and spatially uniform \cite{Serreau2003}
\begin{equation}
	k \frac{dR}{d^3k} = \frac{i}{2 (2\pi)^3} \Pi_{12}(K)^{\,\mu}_{\enspace\mu}
\end{equation}
Here, \(k\) is the photon energy and \(\Pi_{12}\) is one component of the photon polarization tensor in the real-time formalism \cite{Bellac2011}. 

\begin{figure}
\centering
\begin{subfigure}[b]{0.95\columnwidth}
\begin{tabular}{c c c} 
	\includegraphics[width=0.28\linewidth]{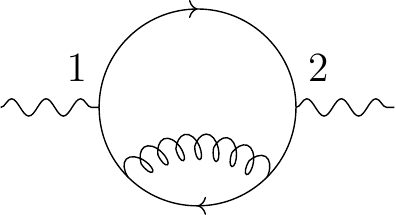}

	\includegraphics[width=0.28\linewidth]{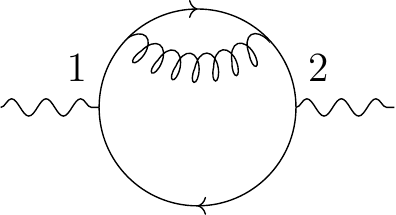}
	
	\includegraphics[width=0.28\linewidth]{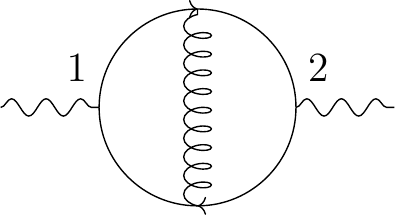}
\end{tabular}
\label{fig:hard_loops} 
\end{subfigure}

\begin{subfigure}[b]{0.95\columnwidth}
\centering
\begin{tabular}{l l}
	\hspace{-0.48cm} 
	\includegraphics[width=0.28\linewidth]{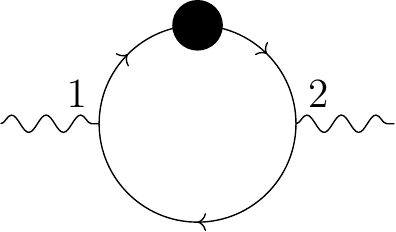}

	\includegraphics[width=0.28\linewidth]{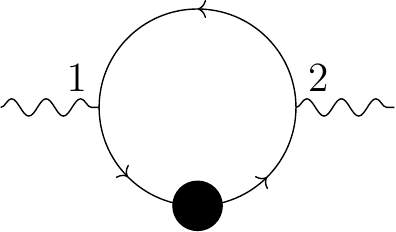}
\end{tabular}
\label{fig:soft_loops} 
\end{subfigure}
\caption{Top row:  Diagrams contributing at leading order for hard quark loops. Bottom row: Diagrams contributing at leading order for hard quark loops.}
\label{photon_selfE}
\end{figure}

The diagrams in the top row of Fig. \ref{photon_selfE} contribute at leading order to the photon polarization tensor when the loop momentum is hard. Using finite-temperature cutting rules these diagrams can be transformed into a kinetic theory equation for the rates \cite{Kapusta1991}
\begin{multline}
k \frac{dR}{d^3 k} = \sum_{\mathrm{channels}} \int_P \, \int_{P^{'}} \, \int_{K^{'}} \frac{1}{2(2\pi)^3}
\\ (2\pi)^4 \delta^{(4)}(P+P'-K-K') \,
\left| \mathcal{M}\right|^2 f(P) \, f(P')\, (1 \pm f(K')).
\end{multline}
Here \(\mathcal{M}\) is the amplitude for two-to-two scattering with a photon in the final state. Additionally, diagrams with soft loop momenta contribute at the same order, see the bottom row of Fig. \ref{photon_selfE}. In order to avoid infrared divergences and include all leading order effects one must use resummed quark propagators for soft momenta. This is done using the method of hard thermal loops (HTL) \cite{Mrowczynski2000}. We included the viscous correction to the resummed propagator. The different structure of hard and soft loop diagrams necessitates a cutoff scale \(q_{\mathrm{cut}}\) between hard and soft loop momenta, such that \(gT \ll q_{\mathrm{cut}} \ll T\). At leading order the results should be independent of \(q_{\mathrm{cut}}\).

\begin{figure}

\begin{subfigure}[b]{1\columnwidth}
\centering
   \includegraphics[width=0.8\linewidth]{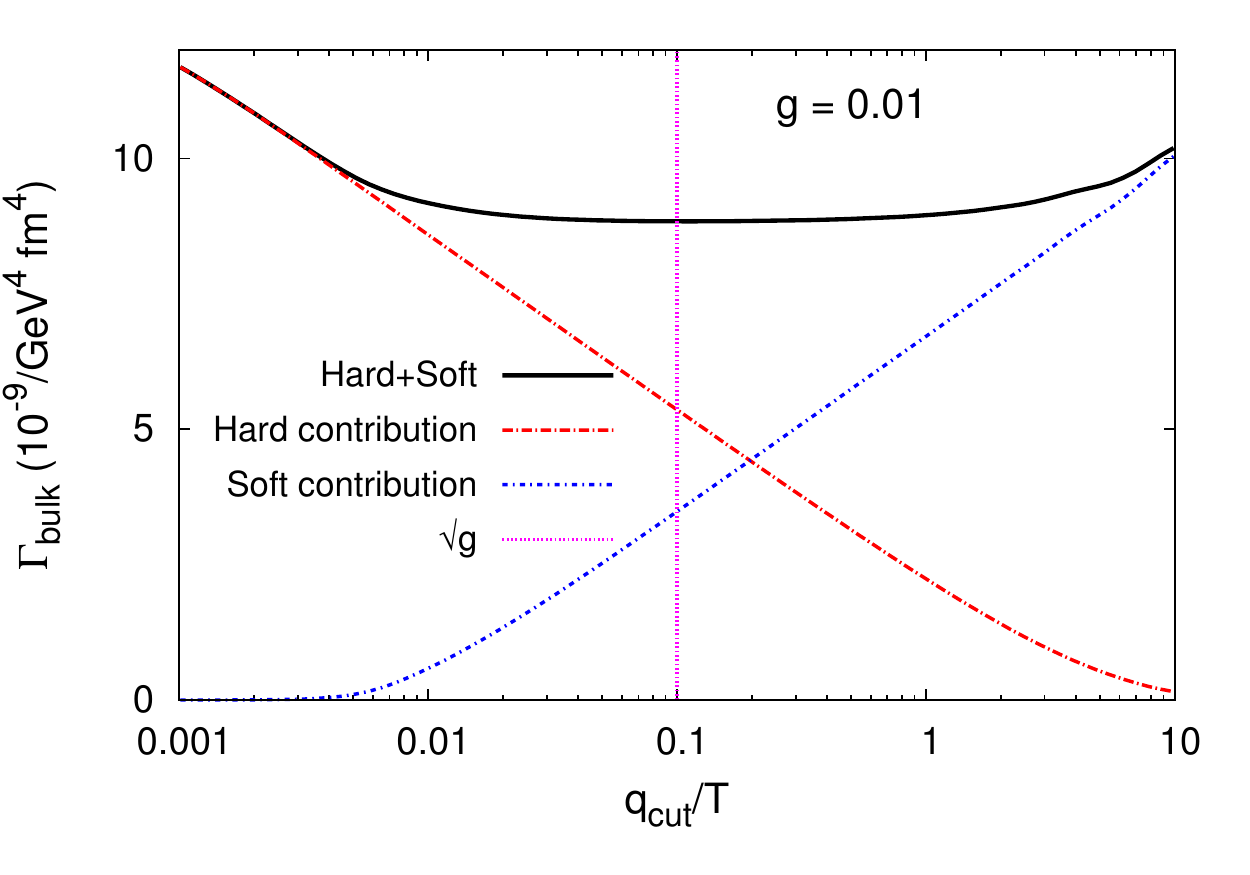}
\end{subfigure}

\begin{subfigure}[b]{1\columnwidth}
\centering
   \includegraphics[width=0.8\linewidth]{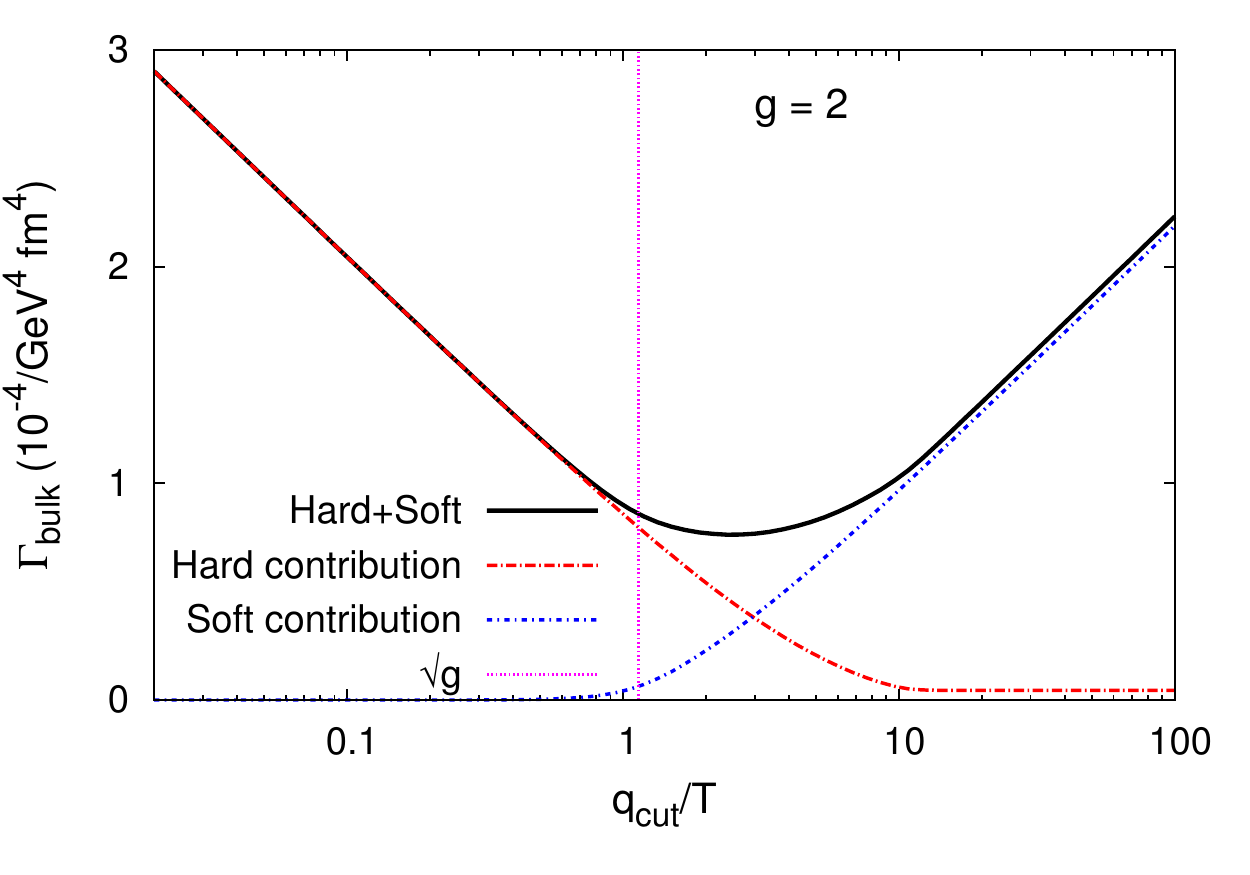}
\end{subfigure}

\caption{Dependence of \(\Gamma_{\mathrm{bulk}}\) on \(q_{\mathrm{cut}}\), the cutoff between hard and soft loops, for two values of the strong coupling constant $g$. }
\label{fig:cut}
\end{figure}

The crucial ingredient in the calculation of viscous corrections is the momentum distribution. 
The equilibrium distribution, \(f_0\), is replaced by 
\begin{equation}
f = f_0 + \delta f_{\mathrm{bulk}}.
\end{equation}
where \(\delta f_{\mathrm{bulk}}\) is the bulk viscous correction derived using kinetic theory: 
\begin{equation}
\delta f_{\mathrm{bulk}} = f_{0} (1 \pm f_{0}) \; (E-\frac{m_{\mathrm{th}}^2}{E}) \;\frac{\Pi}{15 (\epsilon + P) (\frac{1}{3} - c_s^2)}
\end{equation}
where \(E = \sqrt{p^2 +m_{\mathrm{th}}^2}\) and \(\Pi\) is the bulk viscous pressure. 
This equation was obtained in the relaxation time approximation for particles with thermal masses \(m_{\mathrm{th}}\) \cite{Paquet2015}. We calculated rates up to first order in \(\delta f\) evaluating the loop and phase space integrals numerically. For further details see the discussion of shear viscous corrections to photon production using a diagrammatic approach in Ref. \cite{Shen2014}. 

The photon production rate can be written as
\begin{equation}
k \frac{dR}{d^3 k} = T^2 \left(\Gamma_{\mathrm{eq}} + \frac{\Pi}{15(\epsilon + P)(\frac{1}{3}-c_s^2)} \Gamma_{\mathrm{bulk}}\right).
\label{Gamma}
\end{equation}
The  quantities \(\Gamma_{\mathrm{eq}}\) and \(\Gamma_{\mathrm{bulk}}\) contain the contribution of \(f_0\) and \(\delta f_{\mathrm{bulk}}\) respectively to photon production in a static, homogenous brick of QGP. 

Fig. \ref{fig:cut} shows the dependence of \(\Gamma_{\mathrm{bulk}}\) on the cut between soft and hard loops \(q_{\mathrm{cut}}\). At very low values of \(g\) there is a wide range of cuts that give the same value of \(\Gamma_{\mathrm{bulk}}\). At more realistic values for heavy-ion collisions, \(g = 2\), \(\Gamma_{\mathrm{bulk}}\) depends more on the cut which only cancels at leading order. We chose the minimal value of \(\Gamma_{\mathrm{bulk}}\) which occurs at roughly  \(q_{\mathrm{cut}}/T = \sqrt{g}\).

\begin{figure}[h]

\begin{subfigure}[bc]{1\columnwidth}
\centering
   \includegraphics[width=0.85\linewidth]{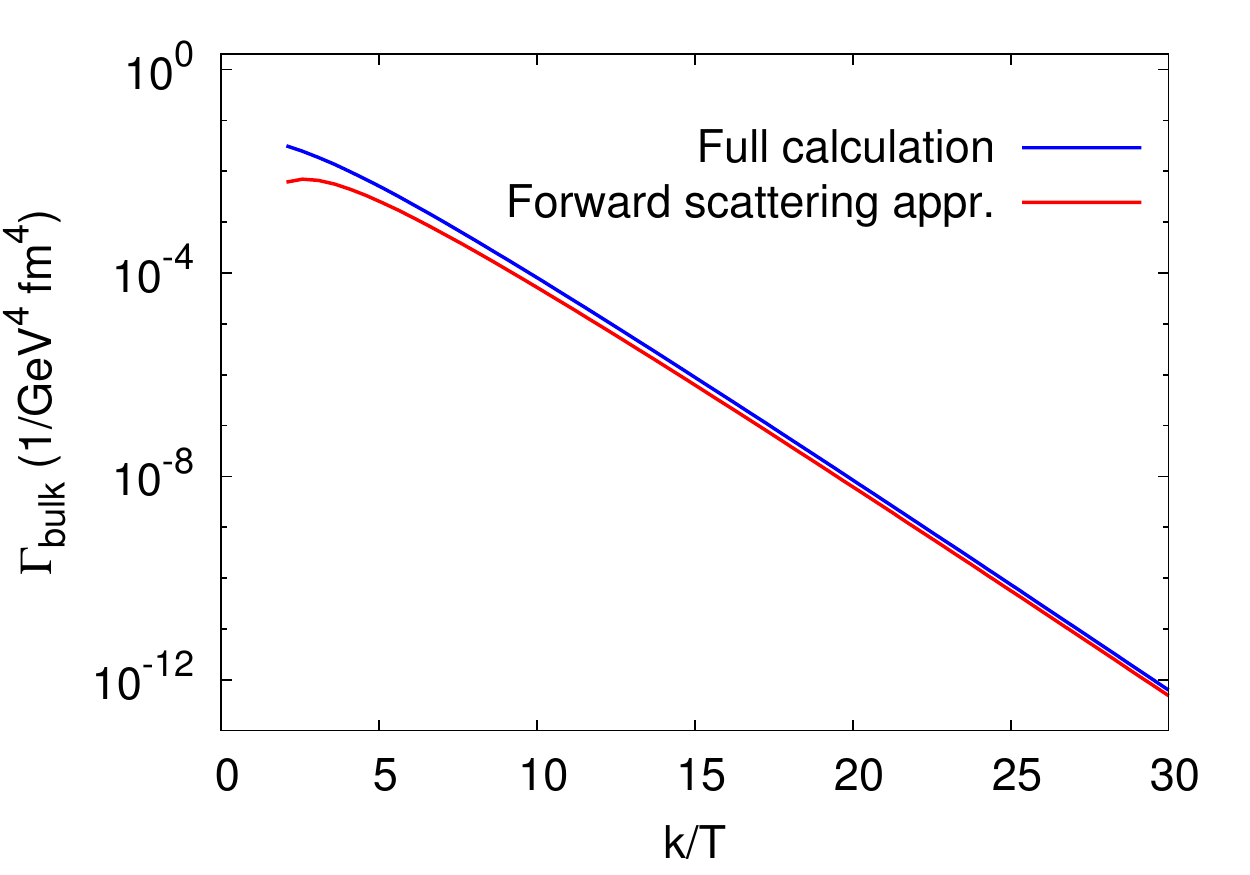}
\end{subfigure}

\centering
\begin{subfigure}[bc]{1\columnwidth}
\centering
\hspace*{0.3cm}
   \includegraphics[width=0.8\linewidth]{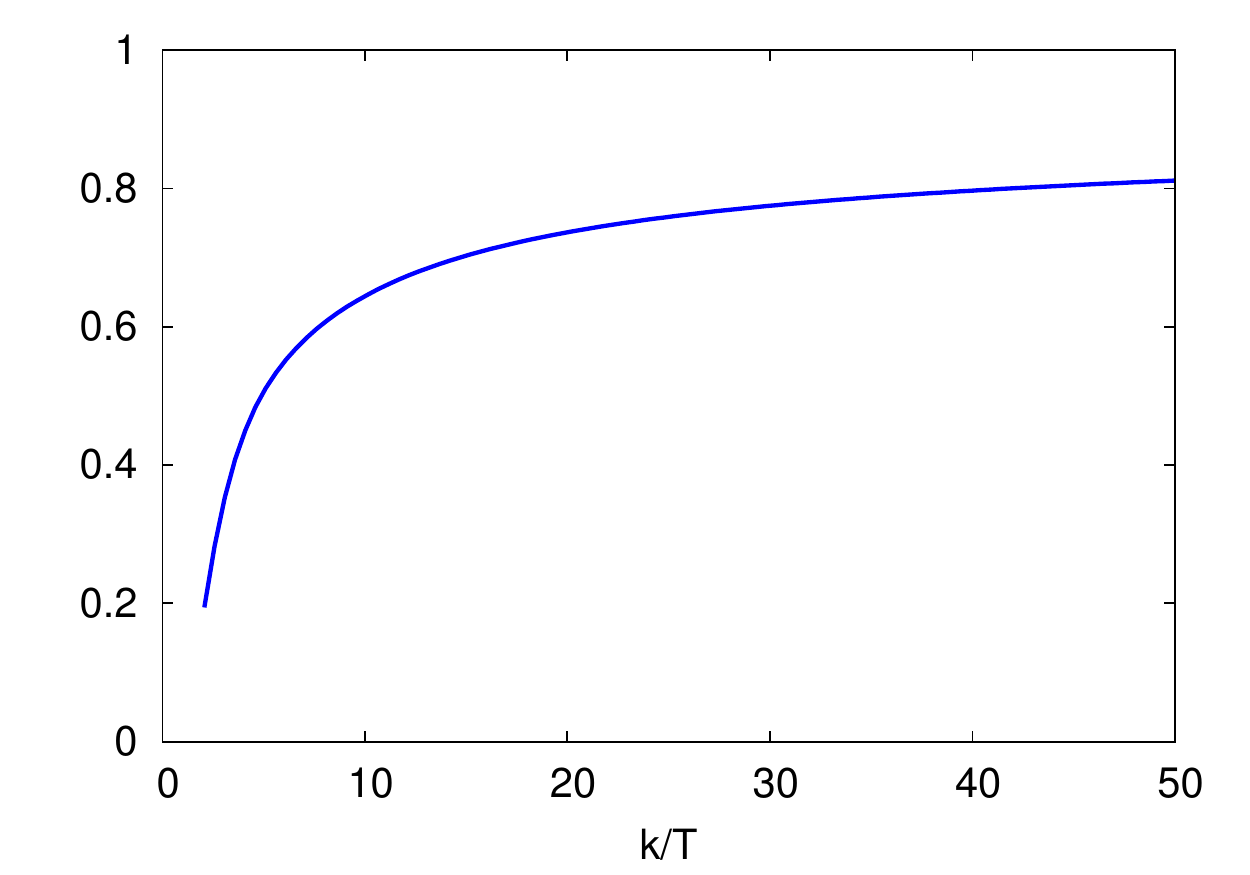}
\end{subfigure}

\caption{Comparison of a full calculation of bulk viscous corrections to photon production and a calculation using the forward scattering approximation. Top panel: Single particle spectrum elements as defined in Eq. (\ref{Gamma}). Bottom panel: Ratio of forward-scattering result to that of the full calculation.}
\label{fig:forwardscattering}
\end{figure}

The full calculation of \(\Gamma_{\mathrm{bulk}}\) can be compared with its value obtained in the forward scattering approximation \cite{Dusling2009}, see Fig. \ref{fig:forwardscattering}. This approximation assumes that the exchanged momentum in the two-to-two scattering diagrams is soft which is valid when the external particles have high momenta. It is furthermore only correct at leading logarithm in \(g_s\) and ignores the viscous correction to the HTL resummation. As expected, the two calculations are similar at high photon momenta but different at lower momenta. 

\section{Hydrodynamic modelling}

Up until now we have only discussed viscous corrections to the photon production rate in a homogeneous brick of QGP. To make contact with experiments we need to integrate these calculations with a hydrodynamic simulation of heavy-ion collisions. 
The spectrum  of thermal photons produced in the medium is then described by 
\begin{equation}
k \frac{dN_{\mathrm{thermal}}}{d^3 k} = \int d^4 x \left[ \left. k \frac{d R}{d^3 k} \left(T(x),E_k \right)\,\right|_{E_k = k \cdot u(x)} \right].
\end{equation}
\(k d R/d^3 k\) only depends on position through hydrodynamic variables such as $u^\mu$, \(T\), and \(\Pi\). For our bulk viscous correction the integral is over all cells in the QGP phase.  The hydrodynamical events were generated with MUSIC \cite{Schenke2010} and used IP-Glasma initial conditions. We looked at Au-Au collisions at \(200\,\mathrm{GeV}\) in a \(0-40\%\) centrality class. For further details see \cite{Paquet2015}. We model all photon production channels as in \cite{Paquet2015}, with the exception of the bulk viscous correction to two-to-two scattering in the QGP phase where the full calculation was used  instead of the forward scattering approximation.

\begin{figure}[h]
\centering
   \includegraphics[width=0.9\linewidth]{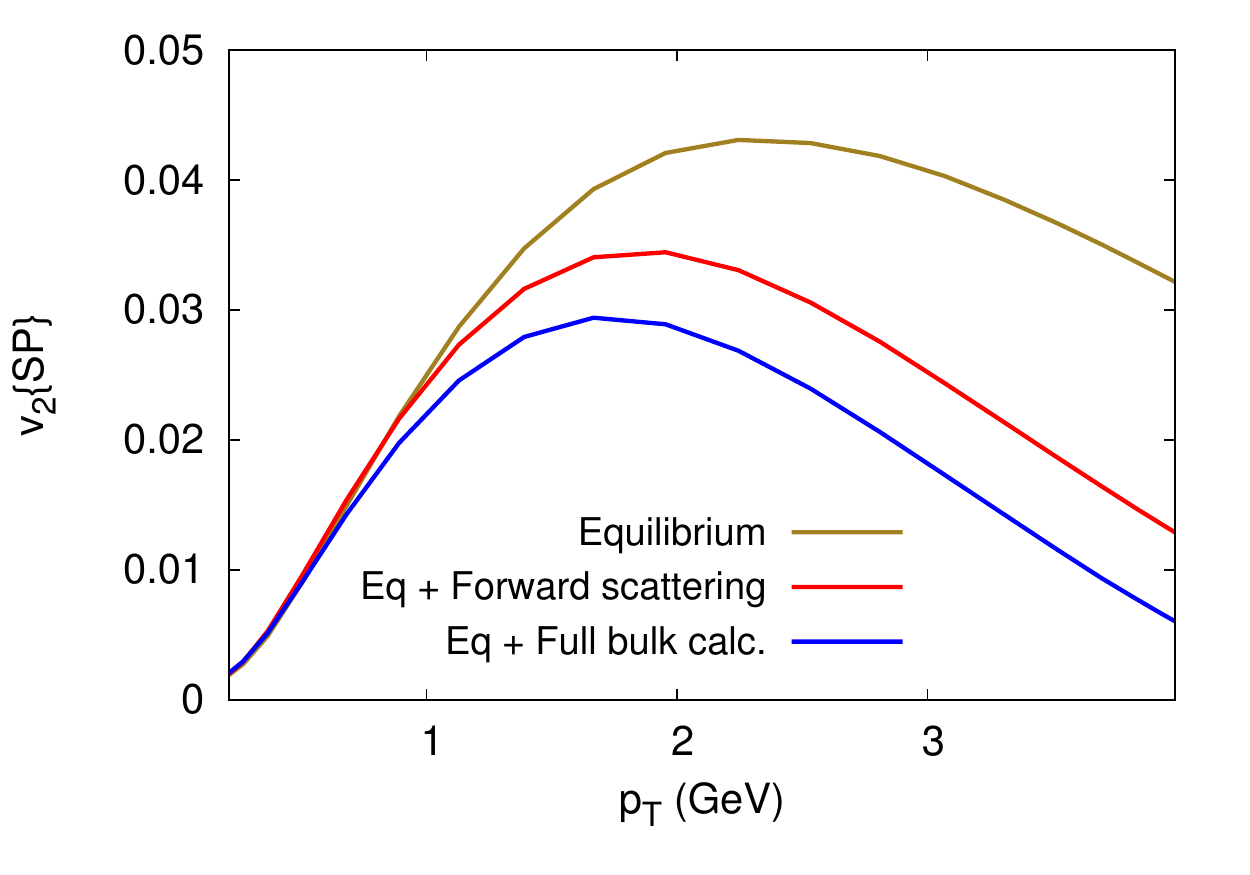}
   \caption{Elliptic flow of thermal photons coming from two-to-two channels in QGP. The curves refer to the treatment of the bulk viscous correction.}
	\label{fig:v22to2}
\end{figure}

\begin{figure}[h]
\centering
   \includegraphics[width=0.9\linewidth]{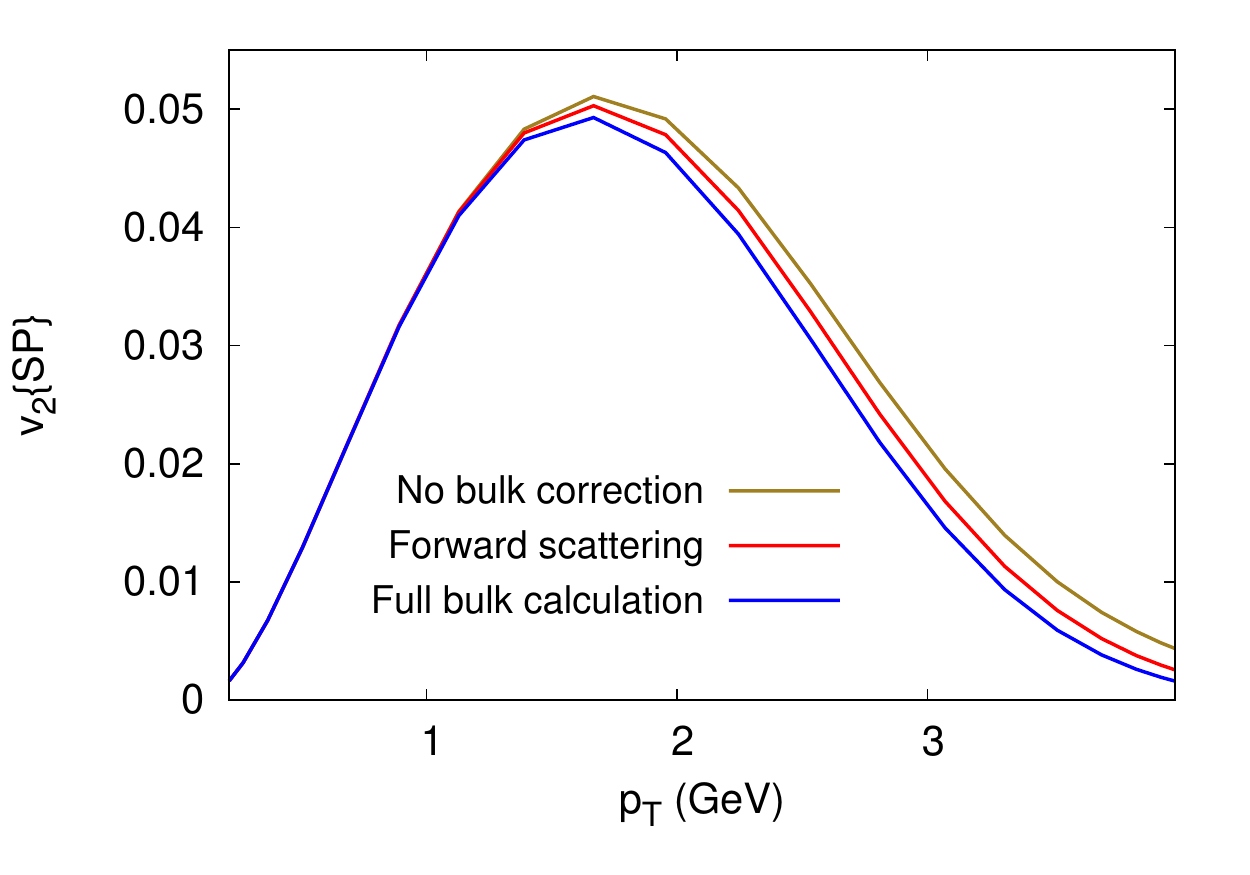}
   	\caption{Elliptic flow of direct photons. The curves refer to  the treatment of the bulk viscous correction to two-to-two scattering in QGP.}
	\label{fig:v2direct}
\end{figure}

%

The bulk viscous correction effect is moderate for the photon spectrum, and becomes more considerable for the elliptic flow. Correcting for bulk viscosity creates a $\sim20\%$ decrease in the yield for photons from $ 2 \to 2$ processes with $p_T\sim2$ GeV. Fig. \ref{fig:v22to2} shows the elliptic flow of photons coming from two-to-two scattering in the QGP phase, and compares results obtained using different bulk viscous corrections to that channel. The underlying hydrodynamical events are the same in all cases and include bulk viscosity. At lower \(p_T\) the bulk viscous correction has little effect but at higher \(p_T\) it reduces the elliptic flow considerably. Fig. \ref{fig:v2direct} shows the elliptic flow of direct photons,  i.e. prompt photons, thermal photons from the hadronic and QGP phase and viscous corrections where they have been calculated. 
The curves only differ in the treatment of the bulk viscous correction to two-to-two scattering in the QGP phase. The difference between the curves is smaller than that in Fig. \ref{fig:v22to2}, owing to the contribution from other channels, but a complete calculation is clearly required. 

\begin{figure}[h]
\centering
   \includegraphics[width=0.8\linewidth]{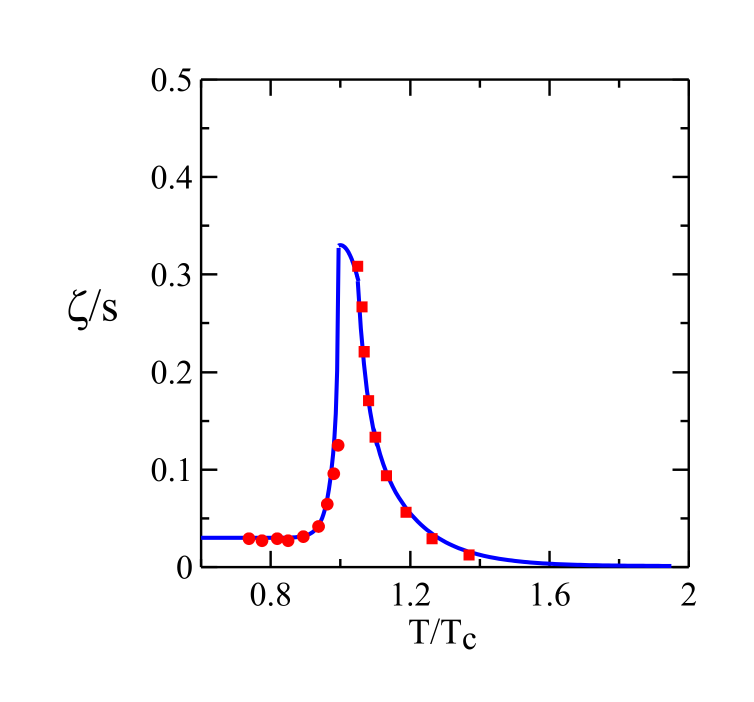}
   	\caption{Bulk viscosity as a function of temperature \cite{Ryu2015}.}
	\label{fig:bulkprofile}
\end{figure}
\begin{figure}[h]
\centering
   \includegraphics[width=0.9\linewidth]{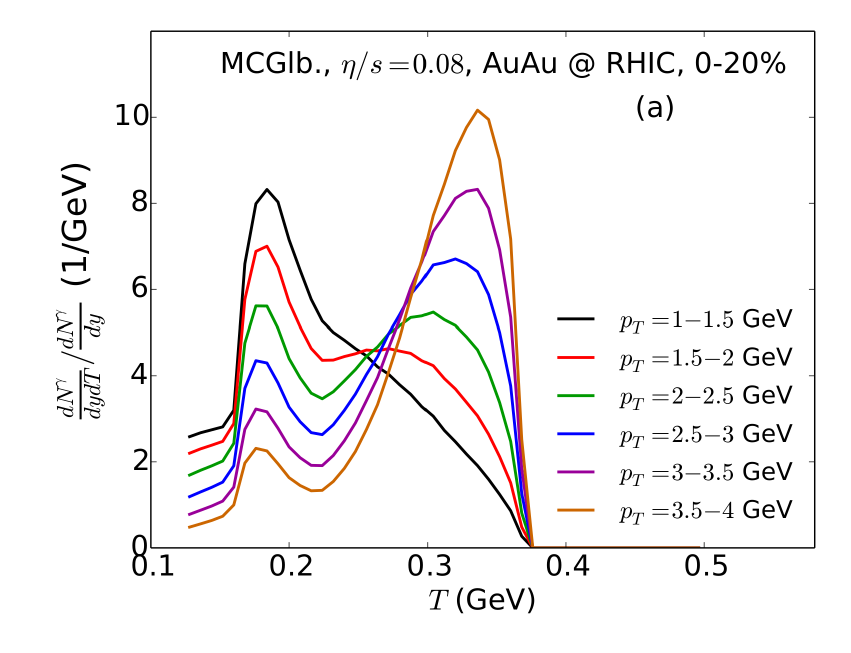}
   	\caption{ Distribution of photon yield in different \(p_t\) bins with the temperature of the emitting fluid cells, for Au Au collisions in the \(0-20\%\) centrality class. This figure is from Ref. \cite{Shen2013}. The events used to make this figure are slightly different from the ones used in this work but the general trend is the same.}
	\label{fig:fromwhichcells}
\end{figure}

The effect of bulk viscous corrections on the photon elliptic flow can be understood in a simple way. Bulk viscosity peaks sharply at a temperature of $T_c = 180$ MeV \cite{Ryu2015}, see Fig. \ref{fig:bulkprofile}. Thus the bulk correction mostly  affects photon production from cells with temperature around \(T_c\). Furthermore \(\delta f_{\mathrm{bulk}}\) is rotationally invariant so elliptic flow of the bulk viscous correction is solely due to the flow of the emitting fluid cell. Fig. \ref{fig:fromwhichcells} shows the yield of photons as a function of temperature and for different windows of photon momentum \cite{Shen2013}. 
Concentrating, for the sake of illustration,  on photons with $p_T \approx 3$ GeV: according to Fig. \ref{fig:fromwhichcells}, those are either blueshifted photons created in low temperature cells with high flow, or photons from high temperature cells with lower flow. 
The bulk correction will suppress emission from the former, which in turn  leads to a net reduced elliptic flow \cite{siggi_prep}. 

\section{Conclusions, outlook, and acknowledgments}
We have presented a calculation of the bulk viscous correction to photon production in QGP through two-to-two channels. This calculation was integrated  with hydrodynamic simulations of heavy-ion collisions. The bulk correction appreciably reduces the elliptic flow at higher photon \(p_T\), while having little effect at lower \(p_T\).

The viscous corrections to all leading-order QGP channels have not been computed so far, as the corresponding Feynman diagrams can have a complicated structure. They can be evaluated for a medium in equilibrium by using the Kubo-Martin-Schwinger relation. In a future publication we will explain  how these diagrams can be evaluated out of equilibrium, thereby allowing the computation of viscous corrections to all leading order channels of photon production in the QGP  in a theoretically consistent fashion \cite{siggi_prep}.

This work was supported in part by the Natural Sciences and Engineering Research Council of Canada, and in part by the DOE under Contract No. DE-SC0012704. C.S. gratefully acknowledges a Goldhaber Distinguished Fellowship from Brookhaven Science Associates, and C. G. gratefully acknowledges support from the Canada Council for the Arts through its Killam Research Fellowship program.

\nocite{*}
\bibliographystyle{elsarticle-num}
\bibliographystyle{unsrt}
\bibliography{jos}

\begin{thebibliography}{10}
\expandafter\ifx\csname url\endcsname\relax
  \def\url#1{\texttt{#1}}\fi
\expandafter\ifx\csname urlprefix\endcsname\relax\def\urlprefix{URL }\fi
\expandafter\ifx\csname href\endcsname\relax
  \def\href#1#2{#2} \def\path#1{#1}\fi

\bibitem{Baier1991}
R.~Baier, H.~Nakkagawa, A.~Niegawa, K.~Redlich, {Production rate of hard
  thermal photons and screening of quark mass singularity}, Z. Phys. C53 (1992)
  433--438.

\bibitem{Kapusta1991}
J.~I. Kapusta, P.~Lichard, D.~Seibert, {High-energy photons from quark - gluon
  plasma versus hot hadronic gas}, Phys. Rev. D44 (1991) 2774--2788, [Erratum:
  Phys. Rev.D47,4171(1993)].

\bibitem{AMY2001}
P.~B. Arnold, G.~D. Moore, L.~G. Yaffe, {Photon emission from ultrarelativistic
  plasmas}, JHEP 11 (2001) 057.

\bibitem{Ryu2015}
S.~Ryu, J.~F. Paquet, C.~Shen, G.~S. Denicol, B.~Schenke, S.~Jeon, C.~Gale,
  {Importance of the Bulk Viscosity of QCD in Ultrarelativistic Heavy-Ion
  Collisions}, Phys. Rev. Lett. 115~(13) (2015) 132301.

\bibitem{Paquet2015}
J.-F. Paquet, C.~Shen, G.~S. Denicol, M.~Luzum, B.~Schenke, S.~Jeon, C.~Gale,
  {Production of photons in relativistic heavy-ion collisions}, Phys. Rev.
  C93~(4) (2016) 044906.

\bibitem{Serreau2003}
J.~Serreau, {Out-of-equilibrium electromagnetic radiation}, JHEP 05 (2004) 078.

\bibitem{Bellac2011}
M.~L. Bellac, {Thermal Field Theory}, Cambridge University Press, 2011.

\bibitem{Mrowczynski2000}
S.~Mrowczynski, M.~H. Thoma, {Hard loop approach to anisotropic systems}, Phys.
  Rev. D62 (2000) 036011.

\bibitem{Shen2014}
C.~Shen, J.-F. Paquet, U.~Heinz, C.~Gale, {Photon Emission from a Momentum
  Anisotropic Quark-Gluon Plasma}, Phys. Rev. C91~(1) (2015) 014908.

\bibitem{Dusling2009}
K.~Dusling, {Photons as a viscometer of heavy ion collisions}, Nucl. Phys. A839
  (2010) 70--77.

\bibitem{Schenke2010}
B.~Schenke, S.~Jeon, C.~Gale, {(3+1)D hydrodynamic simulation of relativistic
  heavy-ion collisions}, Phys. Rev. C82 (2010) 014903.

\bibitem{Shen2013}
C.~Shen, U.~W. Heinz, J.-F. Paquet, C.~Gale, {Thermal photons as a quark-gluon
  plasma thermometer reexamined}, Phys. Rev. C89~(4) (2014) 044910.

\bibitem{siggi_prep}
S.~Hauksson~et al., in preparation.

\end{thebibliography}







\end{document}